\documentclass[aps,prl,reprint,twocolumn,footinbib,showpacs,superscriptaddress,floatfix]{revtex4-1}
\usepackage{graphicx}
\usepackage{dblfloatfix}
\usepackage{amssymb}
\usepackage{amsmath}
\usepackage{times}
\usepackage{bm}

\begin{document}

\newcommand{\cc}{\color{red}} 
	
\title{Faceted patterns and anomalous surface roughening driven by 
long-term correlated noise}

\author{Alejandro Al\'es}
\affiliation{Instituto de Investigaciones F\'isicas de Mar del Plata (IFIMAR),
	CONICET--Universidad Nacional de Mar del Plata, Argentina}

\author{Juan M. L\'opez}
\email{lopez@ifca.unican.es}
\affiliation{Instituto de F\'isica de Cantabria (IFCA), CSIC-Universidad de
	Cantabria, 39005 Santander, Spain}

\date{\today}

\begin{abstract}
We investigate Kardar-Parisi-Zhang (KPZ) surface growth in the presence of 
long-term correlated noise. By means of extensive numerical simulations of 
models in the KPZ universality class we find that, as the noise correlator range 
increases, the surface develops a pattern of macroscopic facets that completely 
dominate the dynamics and induce anomalous kinetic roughening. This novel 
phenomenon is not described by the conventional dynamic renormalisation group 
calculations and can explain the singular behavior observed in the analytical 
treatment of this problem in the seminal paper of 
Medina {\it et al} [Phys. Rev. A {\bf 39}, 3053 (1989)]. 
\end{abstract}

\pacs{Valid PACS appear here}

\maketitle

The dynamics of surfaces and interfaces driven by random fluctuations has many 
applications in modern condensed matter science and statistical physics 
including the description of surfaces formed by particle deposition 
processes in thin-film growth ({\it e.g.} molecular-beam epitaxy, sputtering, 
electrodeposition, and chemical-vapor deposition)~\cite{Barabasi.Stanley1995_book, Krug1997}, 
advancing fracture cracks in disordered materials~\cite{Alava2006}, 
and fluid-flow depinning in disordered media~\cite{Alava.etal_2004}. 
The dynamics of scale-invariant interfaces 
is also relevant to understand many classical problems in statistical mechanics 
including directed polymers in random media, minimal energy paths, and localization in 
disordered media~\cite{Kardar_1998,Halpin-Healy.Zhang_1995}. Remarkably, it has 
also been recently found that there exists 
a deep connection of interface kinetic roughening with the evolution of 
perturbations, the so-called Lyapunov vectors, in chaotic spatially extended 
dynamical systems~\cite{Pazo2013,Pazo2016,Pikovsky.Politi_1998,Pazo2014,Pazo2008,
Pikovsky.Politi_2001,Szendro2007,Pazo2009}.
All these theoretical interconnections between apparently
distant problems make the understanding 
of all aspects of kinetic 
surface roughening a central theme in modern statistical physics.

The time evolution, Langevin-type, equation for the surface height $h(\mathbf{x},t)$ in 
most of the above mentioned systems satisfies a set of fundamental symmetries 
(rotational, translational in $\mathbf{x}$, time invariance, etc), including the 
fundamental {\em shift} symmetry $h \to h + c$, where $c$ is a 
constant~\cite{Barabasi.Stanley1995_book}. The latter immediately leads to 
scale-invariant surfaces/interfaces without fine tuning of any external 
parameters or couplings, which implies generic power-law decay of the 
spatio-temporal surface height correlations~\cite{Hentschel_1994}. 
In a nutshell, a surface is said to be scale invariant 
if its statistical properties remain unchanged after re-scaling of space and 
time according to the transformation 
$h(\mathbf{x},t) \to b^{\alpha} h(b \, \mathbf{x}, b^{1/z} \, t)$, 
for any scaling factor $b>1$ and a certain combination of critical  exponents 
$\alpha$ and $z$~\cite{Barabasi.Stanley1995_book,Krug1997}. 

In kinetic surface roughening the Kardar-Parisi-Zhang (KPZ)~\cite{Kardar.etal_1986} 
equation plays a central role as the simplest, nonlinear, 
out-of-equilibrium model in the 
continuum exhibiting scale-invariant solutions. KPZ describes a 
universality class of surface growth that brings together many different surface 
dynamics that share the same symmetries including Eden growth, ballistic 
deposition, and Kim-Kosterlitz growth models among many 
others~\cite{Barabasi.Stanley1995_book,Krug1997}. The KPZ 
equation describes the evolution of the interface height $h(\mathbf{x},t)$ at 
time $t$ and substrate position $\mathbf{x}$ in $d+1$ dimensions and is 
given by~\cite{Kardar.etal_1986}
\begin{equation}
\partial_t h(\mathbf{x},t)= \nu \mathbf{\nabla}^2 h + 
\lambda (\mathbf{\nabla} h)^2 + \eta(\mathbf{x},t), 
\label{kpz}
\end{equation}
where $\eta(\mathbf{x},t)$ is an uncorrelated noise
\begin{equation}
\langle \eta(\mathbf{x},t) \eta(\mathbf{x}',t') \rangle = 
2D \delta(\mathbf{x}-\mathbf{x}') \delta(t-t').
\end{equation} 
The KPZ equation with uncorrelated noise has scale-invariant solutions as can be 
rigorously proven by calculating height-height correlation functions by means of 
dynamic renormalization group (RG) techniques~\cite{Kardar.etal_1986}. 
In $1+1$ dimensions 
one can obtain the exact critical exponents $\alpha = 1/2$ and $z=3/2$ almost 
straightforwardly (see for instance~\cite{Barabasi.Stanley1995_book}) 
after realizing that, on the one hand, the stationary solution 
of the Fokker-Planck equation associated with the Langevin dynamics, Eq.\ 
(\ref{kpz}), in $d=1$ is identical to that of the linear ($\lambda =0$) case, implying 
$\alpha=1/2$ and, on the other hand, that Eq.\ (\ref{kpz}) satisfies Galilean 
invariance (in any dimension), which implies the hyperscaling relation 
$\alpha + z = 2$. In contrast, exact exponents are not known in higher dimensions,
due to the existence of a strong-coupling fixed point for $d>1$ that cannot be 
approached with pertubative RG calculations,
and have been estimated from 
numerical simulations~\cite{Kim.Kosterlitz_1989,Kim_1991, Moser1991,
Ala-Nissila1992,Ala-Nissila1993,Forrest1990,Miranda2008,Alves2014,Alves2016} 
or analytical approximations of several 
sorts~\cite{Colaiori2001,Canet2010, Schwartz1992,Fogedby2005,Fogedby2006}. 

In many applications the noise is actually correlated. If correlations 
are short-ranged one expects that, in the long wavelength limit, the critical 
exponents should remain exactly the same as in the case of uncorrelated noise. 
Indeed, this result can be rigorously proven by, for instance, RG arguments. 
However, in systems where noise 
correlations are long-ranged or long-termed the critical exponents do depend 
on the noise correlator decay exponents, as was early shown by Medina 
{\it et al.}~\cite{Medina.etal_1989} using perturbative RG.

While the effect of spatially correlated noise in KPZ has been extensively 
studied in the literature~\cite{Halpin-Healy1989,Meakin_1989,Zhang1990,
Margolina1990,Amar1991,Peng1991,Wu1995,Pang1995,Jeong1995,Li1997,Janssen1999,Kuittu1999,
Katzav1999,Verma2000,Kloss2014}, 
the case of algebraic temporal correlations has 
remained virtually unexplored so far~\cite{Lam1992,Katzav2004, 
Strack2015, Song_2016}. 
Interestingly, the effect of long-term 
correlated noise in kinetic roughening has earned renewed interest as it
has been very recently shown to play a crucial, and 
not yet completely understood, role in the interface 
picture of infinitesimal perturbations (Lyapunov vectors)~\cite{Pazo2009,
Romero-Bastida2010,Pazo2014} and the scaling of the Lyapunov exponent fluctuations in 
spatio-temporal chaos~\cite{Pazo2016}.

In this Letter we find that KPZ growth in the presence of temporally 
correlated noise gives rise to surfaces with a faceted structure and the 
accompanying anomalous kinetic roughening. The origin of the faceted 
pattern can be traced back to
the strong localization properties of the field 
$\phi(\mathbf{x},t) = \exp h(\mathbf{x},t)$ and the same 
mechanism is expected to be relevant for other surface growth models. 
Our conclusions are based upon 
extensive numerical simulations of two models in 
the same universality class. We studied ballistic deposition as an
example of a discrete growth model with KPZ symmetries. 
We also performed a direct numerical integration of 
the KPZ equation (\ref{kpz}) with a long-term correlated 
noise.

{\paragraph{Models.-} In our numerical study we needed to
generate very long time series of random numbers 
with long temporal correlations at every spatial position $x$. In particular, 
one must generate a spatially uncorrelated time series $\eta(x,t)$ at every 
lattice site $x$ and be certain that the power spectrum $\langle|\hat 
\eta(x,\omega)|^2\rangle$, where $\langle \centerdot \rangle$ 
denotes noise average, exhibits excellent scaling at low frequencies 
so that, at leading order, 
$\langle|\hat \eta(x,\omega)|^2\rangle \sim 
\omega^{-2\theta}$ for $\omega \to 0$ at every lattice site $x$ and 
the noise correlator scales as
\begin{equation}
\langle \eta(x,t) \eta(x',t')\rangle = 2D \; \delta_{x,x'} \; |t-t'|^{2\theta-1},
\label{noise}
\end{equation}
for time differences $|t-t'| > L^z$. In this way, we can be 
certain that we have spatially uncorrelated noise with power-law scaling of 
the temporal  correlations for times differences up to, at least, 
the saturation time in a system of size $L$. The exponent $\theta \in [0,1/2)$ 
characterizes the temporal correlation range of the noise that becomes
more long-term correlated as $\theta$ is increased from zero. 

For each lattice site $x$ we generated a noise sequence 
parametrized by $t$ using the Mandelbrot's fast fractional Gaussian 
noise generator \cite{Mandelbrot1969, Mandelbrot1971}. 
This algorithm produces a random sequence of 
Gaussian distributed numbers $Z (t)$ which can be used in the numerical 
integration of the KPZ equation, $\eta(x,t) \equiv Z_x(t)$, by generating 
$L$ independent Mandelbrot sequences $Z_1(t), Z_2(t) \cdots, Z_L(t)$. In the case of 
the simulations of particle deposition by ballistic deposition 
we map $\eta(x,t) = 1$ if $Z_x(t) > 0$ and $\eta(x,t) = 0$ otherwise. This 
{\em digitalization} of the noise enhances 
the statistics of the simulations by avoiding the formation of overhangs on the 
surface~\cite{Lam1992}. We checked that the noise generated has a 
correlator with the correct scaling at very long times with the desired decay 
exponent $\theta$ (see Supplemental Material~\cite{SM}).

We simulated ballistic deposition by implementing the following discrete 
time evolution for the surface: 
\begin{equation*}
h(x,t+1) = \mathrm{Max} [h(x,t) + \eta(x,t), h(x-1,t), h(x+1,t)]
\end{equation*}
where the height $h(x,t)$ is an integer and the noise $\eta \in \{0,1\}$ is temporally 
correlated as in Eq.~(\ref{noise}) with an exponent $0 \leq \theta < 1/2$. 
Periodic boundary conditions were used and the algorithm is updated in parallel 
so that growth is attempted at all even (odd) sites at even (odd) time steps.

We also carried out a numerical integration of the KPZ equation  
with temporally correlated noise. For reasons that will become clear later 
the noise correlator (\ref{noise}) yields surfaces 
that develop a faceted pattern with an increasing 
slope, $\langle\vert\nabla h\vert\rangle$, as the correlation 
exponent $\theta$ is increased above certain threshold. 
This leads to a numerical instability in finite time for any discretization of 
Eq.~(\ref{kpz}). To avoid this we replace the nonlinear term $\lambda (\nabla 
h)^2$ by an arbitrary function $\lambda f[(\nabla h)^2]$ that saturates 
for large values of 
the argument. We have checked several choices for the control function $f(y)$ 
with similar results. For definiteness, the numerical results presented below 
correspond to the choice $f(y) = (1-e^{-cy})/c$, where $c>0$ is a constant. We used 
$c = 0.1$ in our simulations.
This is equivalent to include the infinite series of 
nonlinear terms 
$\lambda (\nabla h)^2 [1 + \sum_{n=1}^\infty (-c)^{n} (\nabla h)^{2n}/(n+1)!]$ 
in the evolution equation, Eq.~(\ref{kpz}), while respecting all KPZ symmetries. 
This trick stabilizes the numerical scheme for any $0 < c < 1$, as 
occurs in other growth models in which the average local slope, 
$\langle\vert\nabla h\vert\rangle$, becomes large~\cite{Dasgupta1996, Dasgupta1997}. 
We discretized Eq.~(\ref{kpz}), with $f[(\nabla h)^2]$
replacing $(\nabla h)^2$, with an Euler finite-differences 
scheme and the noise correlator given in Eq.~(\ref{noise}) 
(see Supplemental Material~\cite{SM}).

\begin{figure}
	\centerline{\includegraphics[width=80mm,clip]{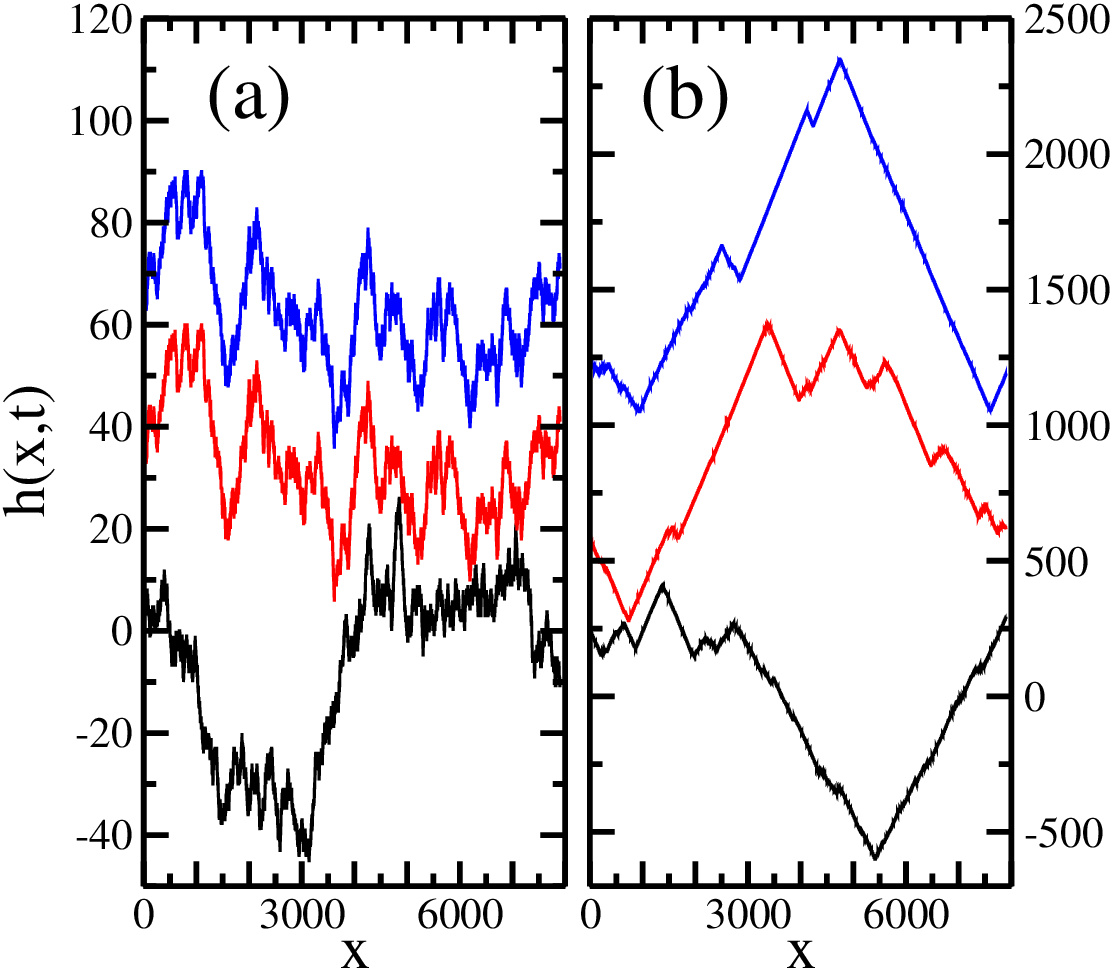}}
	\caption{Snapshots at three different times, $t_1=4 \times 10^4$(black), 
	$t_2=7 \times 10^4$(red), and $t_3=1 \times 10^5$ (blue) time steps, 
	of the ballistic deposition model for the noise correlator
	exponents $\theta=0.15$ (a) and $\theta=0.47$ (b), in a system of size $L=8192$.
	Profiles were shifted in the vertical axis for easy view.}
	\label{fig:fig1}
\end{figure}

\paragraph{Numerical Results.-} In all our simulations the surface is 
started from a initially flat profile $h(x,0) = 0$ and periodic boundary 
conditions are used. As time evolves the surface becomes progressively 
rough and height fluctuations grow under the action of the random noise. 
Surface correlations are measured by means of height-height correlations,
surface width, and the structure factor at different times in the evolution. 
Surface fluctuations saturate and become stationary at a characteristic time
that scales with the system size, $t_\times \sim L^z$. 

In Fig.~\ref{fig:fig1} we plot the 
surface height for the ballistic deposition simulation in a 
system of size $L=8192$ and correlation exponents $\theta = 0.15$ and 
$\theta = 0.47$. It becomes apparent the spontaneous formation of a faceted 
pattern as the correlation index is increased. We found similar pattern formation 
in the numerical integration of the KPZ equation 
(see Supplemental Material~\cite{SM}). 

Apart from the evident change in the visual aspect of the height profiles 
for $\theta > \theta_\mathrm{th}$, the coexistence of faceted patterns with 
scale invariant dynamics also leads to important effects in 
the scaling behavior of the surface height correlations. According to 
Ramasco {\it et al.}~\cite{Ramasco.etal_2000} theory of 
generic kinetic roughening, scale-invariant faceted surfaces obey 
inherently different scaling functions (and exponents) arising from 
the patterned structure. Following Ramasco {\it et al.}~\cite{Ramasco.etal_2000} 
the most general description of the scaling properties of a growing 
surface is best achieved by using the structure factor 
$S(k,t) = \langle \widehat{h}(\mathbf{k},t) \widehat{h}(-\mathbf{k},t) \rangle$, 
where $\widehat{h}(\mathbf{k},t) \equiv (1/L)^{d/2} 
\int d\mathbf{x} \; h(\mathbf{x},t) \exp(-i \mathbf{k} \cdot \mathbf{x})$ 
is the Fourier transform of the surface height $h(\mathbf{x},t)$, 
and $k = \vert \mathbf{k} \vert$. For kinetically roughening 
surfaces in $d+1$ dimensions we expect 
\begin{equation}
S(k,t) = k^{-(2\alpha + d)} s(kt^{1/z}),
\label{S}
\end{equation}
where the most general scaling function, consistent with scale-invariant dynamics, 
is given by~\cite{Ramasco.etal_2000}
\begin{equation}
s(u) \sim  
\left\{ \begin{array}{lcl}
u^{2(\alpha-\alpha_s}) & {\rm if} &  u \gg 1\\
u^{2\alpha + d} & {\rm if} &  u \ll 1
\end{array}
\right. ,
\label{generic}
\end{equation}
with $\alpha$ being the {\em global} roughness exponent and $\alpha_s$ 
the so-called {\em spectral} roughness exponent~\cite{Ramasco.etal_2000}. 
Standard scaling corresponds to $\alpha_s = \alpha < 1$. However, other situations
may be described within the generic scaling framework, 
including super-roughening and intrinsic anomalous scaling, depending
on the values of $\alpha_s$ and $\alpha$~\cite{Ramasco.etal_2000}.
For faceted surfaces, the case of interest for us here, 
one has $\alpha_s > 1$ and $\alpha \neq \alpha_s$ so that two independent
roughening exponents are actually needed to completely describe the scaling 
properties of the surface~\cite{Ramasco.etal_2000}.

\begin{figure}
	\centerline{\includegraphics[width=75mm,clip]{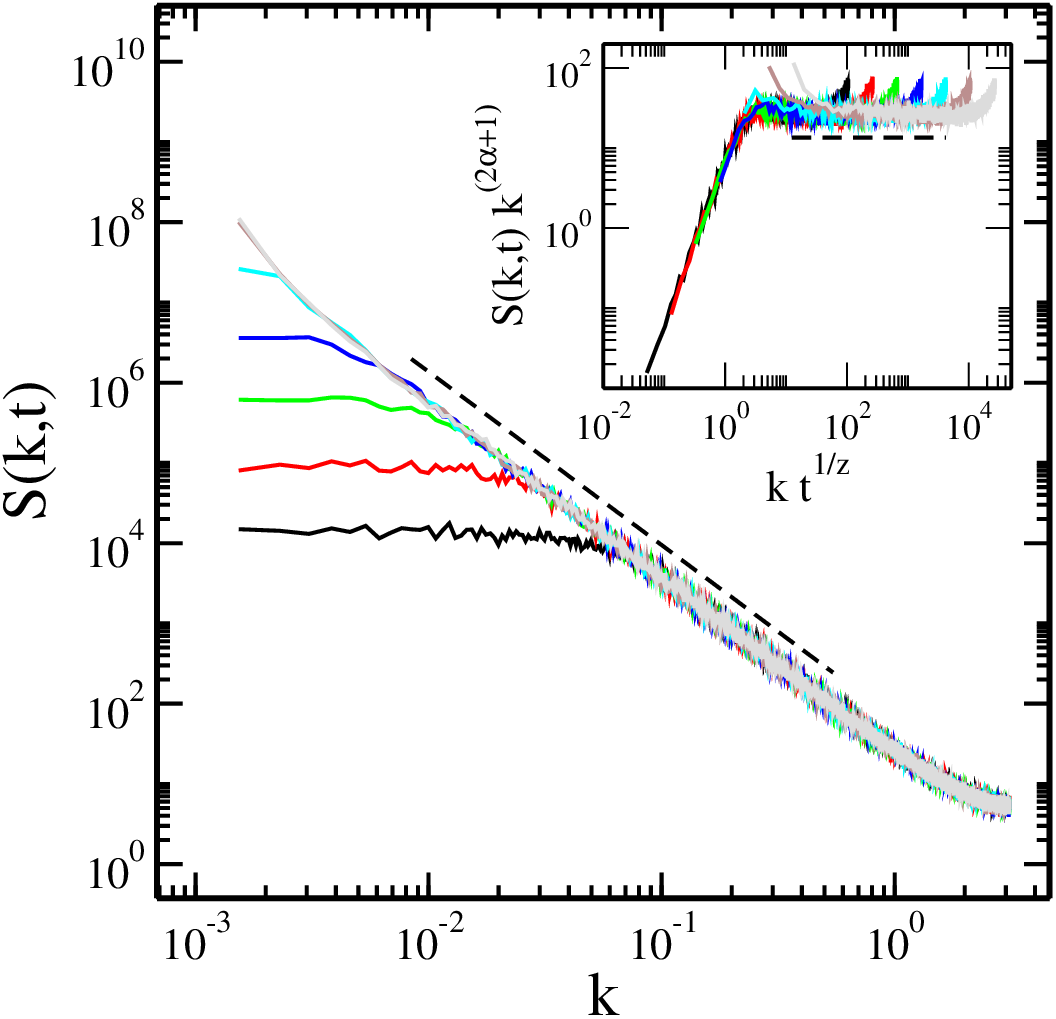}}
	\caption{Structure factor for the ballistic deposition model at different times 
	in a system of size $L=8192$ for $\theta=0.15$. 
	In the inset, we plot the collapse of spectral densities using 
	the critical exponents $\alpha=0.56$ and $z=1.50$. Data were averaged over $100$ 
	independent noise realizations.}
	\label{fig:fig2}
\end{figure}

It is important to remark that only the structure factor $S(k,t)$ allows 
one to obtain the distinctively characteristic spectral roughness exponent 
$\alpha_s$ typical of faceted growing surfaces. For instance, the usual global
surface width $W(L,t) = \langle [h(x,t) - 
\overline{h}(t)]^2 \rangle^{1/2} $, 
where $\overline{h}(t)$ is the spatial average height at time $t$ and the 
brackets $\langle \cdots \rangle$ denote average over noise, 
can be analytically calculated from Eqs.~(\ref{S}) and (\ref{generic}) using 
$W^2(L,t) =  \int \frac{dk}{2\pi} S(k,t)$ to obtain
$W(L,t) = t^{\alpha/z}{\cal F}(L/t^{1/z})$,
with the standard scaling 
function is ${\cal F}(u) \sim u^{\alpha}$ for $u \ll 1$ and ${\cal F}(u) \sim 
\mathrm{const}$ for $u \gg 1$, as was shown in Ref.~\cite{Ramasco.etal_2000}. 
Hence, the anomalous spectral exponent $\alpha_s$ leaves 
no trace in the usual height-height correlation functions.

\begin{figure}
	\centerline{\includegraphics[width=75mm,clip]{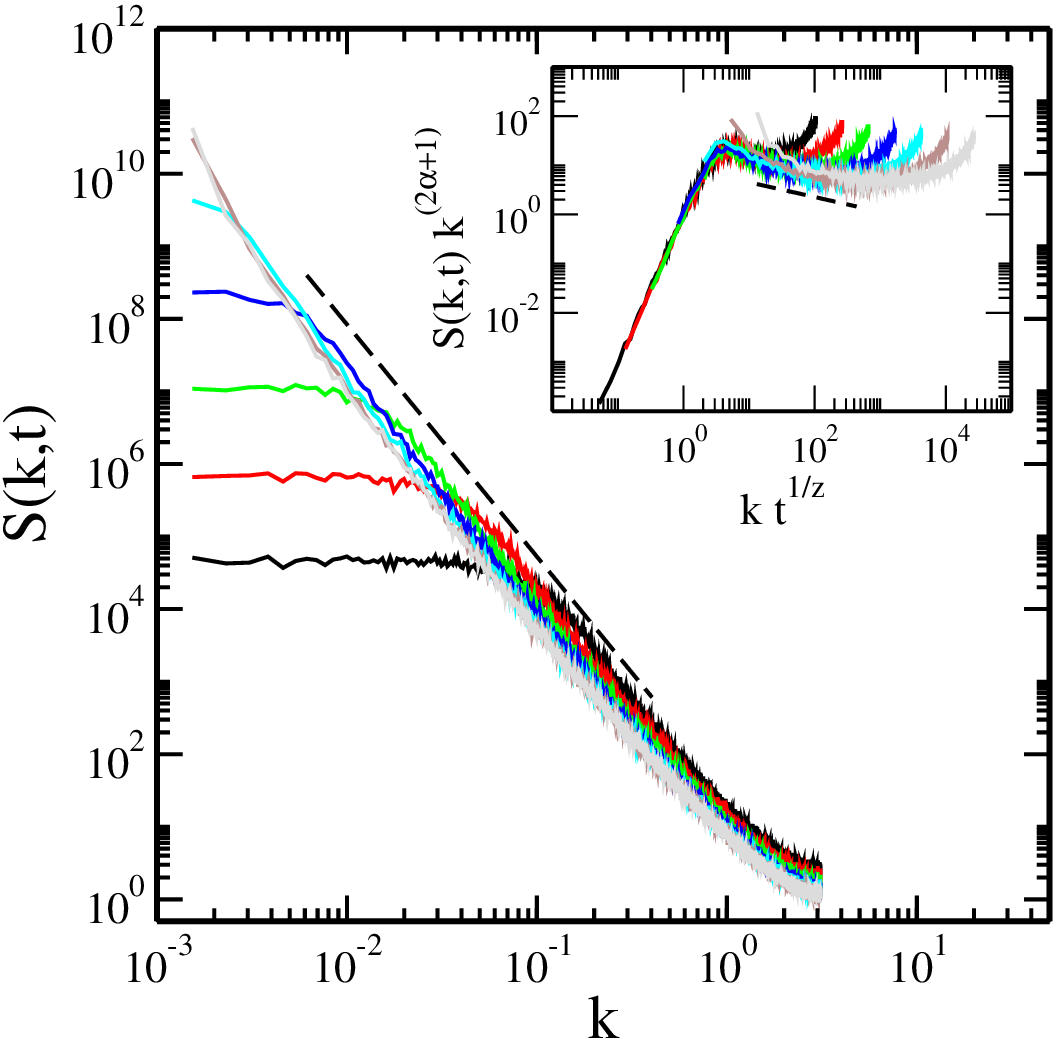}}
	\caption{Structure factor for the ballistic deposition model at different times 
		in a system of size $L=8192$ in the faceted face, for $\theta=0.47$. 
		In the inset, we plot the collapse of spectral densities using 
		the critical exponents $\alpha=1.02$ and $z=1.50$. Data were averaged over $100$ 
		independent noise realizations.}
	\label{fig:fig3}
\end{figure}

In our simulations we analyzed the structure factor $S(k,t)$ of surfaces produced
with the ballistic deposition algorithm and the discretized KPZ equation
for noise with correlation index $\theta$ varying in the interval $[0,1/2)$. 
In Figs.~\ref{fig:fig2} and \ref{fig:fig3} we plot our results for ballistic growth 
with correlated noise exponent $\theta = 0.15$ and $\theta = 0.47$, respectively. These 
$\theta$ values correspond to those plotted in Fig.~\ref{fig:fig1}
for easy comparison. Data collapse analysis was used to obtain 
the scaling behavior of $S(k,t)$ (insets of Figs.~\ref{fig:fig2} 
and \ref{fig:fig3}). This analysis reveals that the structure factor indeed 
exhibits anomalous scaling, as corresponds to faceted scale-invariant 
surface roughening, in the case $\theta = 0.47$ with a spectral roughness 
exponent $\alpha_s = 1.23 \pm 0.02$ and roughness exponent $\alpha = 1.02 \pm 0.01$, 
while $\alpha_s = \alpha = 0.56 \pm 0.02$ ({\it i.e.} standard scaling) for $\theta = 0.15$. 

We have systematically analyzed the scaling behavior of the structure factor 
for the ballistic deposition model and the discretized KPZ equation
as the correlation range of the noise, $\theta$, is varied in the interval $[0,1/2)$.
We computed the scaling exponents $\alpha$ and $\alpha_s$ 
from data collapse analysis of $S(k,t)$ for system sizes 
$L= 2048, 4096$, and $8192$. Note that large system sizes are required for the
surface to develop a faceted pattern before it saturates.
Our main results are summarized in Fig,~\ref{fig:fig4}, where the global roughness 
exponent $\alpha$ and the spectral roughness exponent $\alpha_s$ are plotted 
as a function of the noise correlation index $\theta$
for ballistic deposition growth. Similar results were
obtained for the numerical integration of the KPZ equation with correlated noise
(see Supplemental Material), demonstrating that our 
findings are robust within the universality class. 
Figure~\ref{fig:fig4} clearly shows that 
the roughness exponents split up at $\theta_\mathrm{th} = 0.25 \pm 0.03$ for 
the ballistic deposition model (a similar threshold value
$\theta_\mathrm{th} \approx 0.23$ was found for the KPZ equation~\cite{SM}). 
So that, below the threshold one finds standard scaling 
with $\alpha = \alpha_s$ and no facets.
In contrast, as the noise correlation range is increased above the threshold, 
faceted surfaces are formed and $\alpha_s$ 
moves away from $\alpha$. 
For the sake of comparison, Fig.~\ref{fig:fig4} also shows the main theoretical 
approximations to KPZ with temporally correlated noise. We can see that, while the
global roughness exponent is nicely predicted by the dynamic RG calculations of 
Medina {\it et al.}~\cite{Medina.etal_1989} above the threshold, 
it fails to describe the correlation effects below $\theta_\mathrm{th}$. 
Obviously, no present theory is capable of 
predicting the existence of facets and the associated spectral roughness exponent
as the noise correlation range is increased. 

\begin{figure}
	\centerline{\includegraphics[width=75mm,clip]{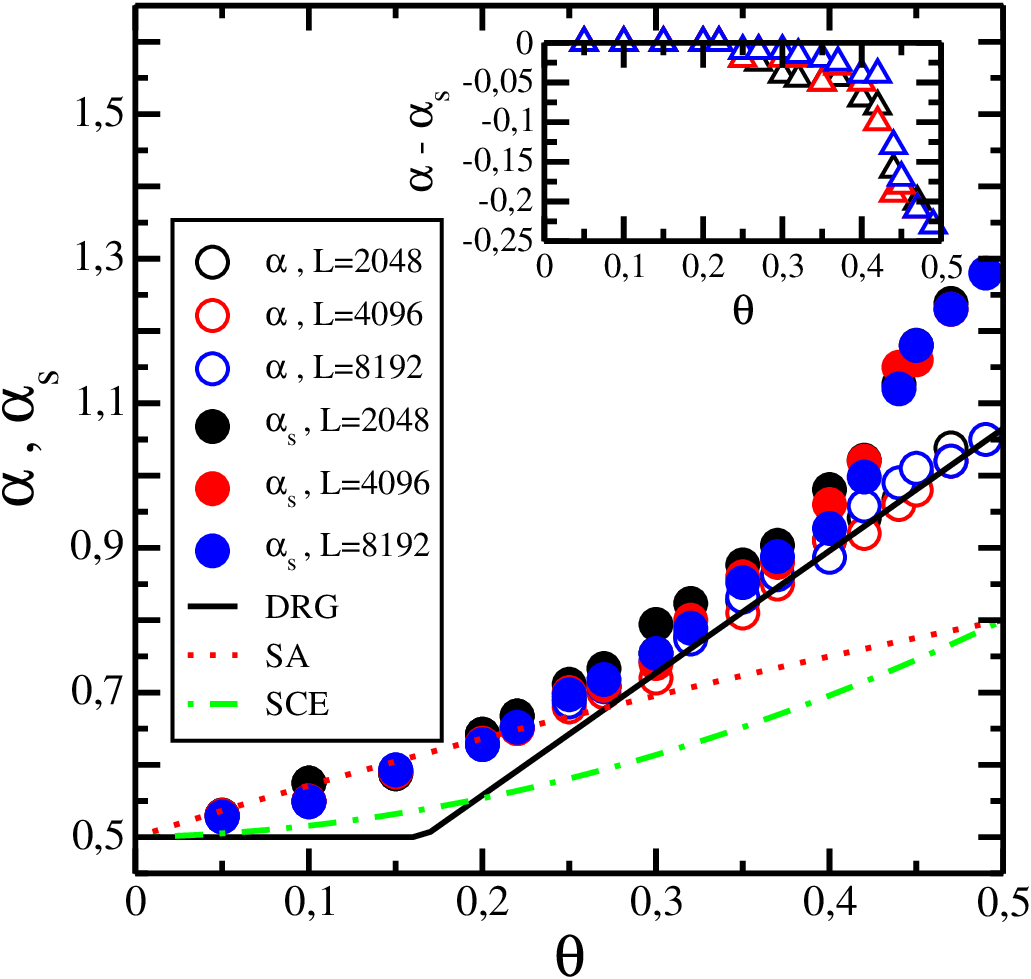}}
	\caption{Global and spectral roughness 
	exponents as a function of the noise correlator 
	index $\theta$ for the ballistic deposition model. 
	For comparison, existing theoretical predictions 
	for $\alpha$ of the RG 
	treatment~\cite{Medina.etal_1989}, Flory scaling approximation 
	(SA)~\cite{Hanfei1993}, and self-consistent expansion approach 
	(SCE)~\cite{Katzav2004} are plotted. The inset shows
	the difference $\alpha - \alpha_s$ as a function of $\theta$ 
	showing the splitting of the exponents at $\theta_\mathrm{th} \approx 0.25$.}
	\label{fig:fig4}
\end{figure}
 
\paragraph{Discussion.-} While we lack of a complete theory to explain 
the numerical findings reported above, we can put forward some arguments to
rationalize the emergence of the faceted patterns and the associated
anomalous scaling.

Let us start by considering the limit case of KPZ dynamics with 
the noise at site each site $\mathbf{x}$ fixed at all times, 
but uncorrelated from site to site-- namely, 
the problem of KPZ with columnar noise:
\begin{equation}
\partial_t h(\mathbf{x},t)= \mathbf{\nabla}^2 h + 
\lambda (\mathbf{\nabla} h)^2 + \eta(\mathbf{x}), 
\label{columnar_KPZ}
\end{equation}
where $\eta(\mathbf{x)}$ is a spatially uncorrelated noise
\begin{equation*}
\langle \eta(\mathbf{x}) \eta(\mathbf{x}') \rangle = 
\delta(\mathbf{x}-\mathbf{x}').
\end{equation*}
This problem corresponds to the limit $\theta = 1/2$ of the long-term 
correlated noise case in Eq.~(\ref{noise}). 
It is well known~\cite{Szendro2007a} that the {\em columnar} KPZ 
equation~(\ref{columnar_KPZ}) exhibits facet formation 
arising from the exponential
localization of the field $\phi(\mathbf{x},t) \equiv \exp h(\mathbf{x},t)$ 
around some random centers $\mathbf{x}_c$. 
The auxiliary $\phi$ can be interpreted 
as the probability density of particles diffusing in a random potential 
$\eta(\mathbf{x})$:
\begin{equation}
\partial_t \phi = \mathbf{\nabla}^2 \phi + \eta(\mathbf{x}) \phi(\mathbf{x},t).
\label{multiplicative}
\end{equation}
The multiplicative-noise term in
Eq.~(\ref{multiplicative}) leads to sharply localized solutions around 
random localization centers. The stochastic field $\phi$ 
has an exponential profile,
$\sim \exp(-\vert \mathbf{x} - \mathbf{x}_c \vert/\xi)$,
around any typical center $\mathbf{x}_c$
with a certain localization extent $\xi$. The solutions of 
Eq.~(\ref{multiplicative}) are, therefore, a superposition
of these exponentially localized functions.   
In turn, this leads to a surface $h$ formed by facets with their cusps 
at the localization centers, 
$h \sim \ln \phi \sim \pm \vert \mathbf{x} - \mathbf{x}_c \vert/\xi$, 
as shown by Szendro {\it et al.}~\cite{Szendro2007a}. There, the reported 
values of the global and spectral roughness exponents were $\alpha = 1.07 \pm 0.05$ 
and $\alpha_s = 1.5 \pm 0.05$, respectively, in $d=1$. 

We conjecture that this localization picture can be 
essentially extended for $\theta < 1/2$. 
Our numerical results indicate that the mechanism for the 
formation of facets based on localization remains valid while 
$\theta > \theta_\mathrm{th}$, where $\theta_\mathrm{th} \approx 1/4$. 
Remarkably, the value $\theta = 1/4$ was already shown to play a special 
role in the perturbative RG approximation of 
Medina {\it et al.}~\cite{Medina.etal_1989}, 
as the point at which the renormalized noise amplitude 
$D^*(\omega)$ has a singular correction at leading $\omega$-order. Further
singularities appear at larger values of $\theta$ making the RG 
treatment ill-constructed.
Our results strongly suggest that the infinitely many singularities 
are a reflection of the failure of the RG treatment to describe the 
appearance of a new exponent $\alpha_s \neq \alpha$. 
Given these results, it becomes clear that a generalization 
of the RG theory would be required to describe the generic 
scaling form~(\ref{generic}) of the spectral function. Such 
a generalization should be able not only to fix the nonphysical 
singularities but it would also provide a coherent 
mathematical picture of anomalous kinetic roughening
as a whole. 

\acknowledgments
J.\ M.\ L.\ thanks D. Paz\'o for discussions and a critical reading of the manuscript.
This work has been partially supported by the Program for 
Scientific Cooperation “I-COOP+” from Consejo Superior de 
Investigaciones Cient\'ificas (Spain) through project No. COOPA20187.
A.\ A. is grateful for the financial support from Programa de 
Pasant\'ias de la Universidad de Cantabria in 2017 and 2018 
(Projects No. 70-ZCE3-226.90 and 62-VCES-648), and CONICET 
(Argentina). J.\ M.\ L is supported by Direcci\'on General 
de Investigaci\'on Cientı\'ifica y T\'ecnica, MICINN (Spain), 
through the project No. FIS2016-74957-P.


%

\end{document}